\documentclass[onecolumn,showpacs,amsmath,amssymb,pra,floatfix]{revtex4}
\usepackage{graphicx}% Include figure files
\usepackage{bm}% bold math
\usepackage{amssymb}
\begin{document}

\title{A time lens for high resolution neutron time of flight spectrometers}
\author{K. Baumann}
\affiliation{Technische Universit\"{a}t, D-80333 M\"{u}nchen,
Germany}
\author{R. G\"{a}hler}
\author{P. Grigoriev}
\altaffiliation[Also at: ]{L.D. Landau Institute for Theoretical
Physics, RAS, Moscow, Russia} \author{E.I. Kats$^*$}
\affiliation{Institut Laue Langevin, F-38042 Grenoble, France }

\begin{abstract}
We examine in analytic and numeric ways the imaging effects of
temporal neutron lenses created by traveling magnetic fields. For
fields of parabolic shape we derive the imaging equations,
investigate the time-magnification, the evolution of the phase space
element, the gain factor and the effect of finite beam size. The
main aberration effects are calculated numerically. The system is
technologically feasible and should convert neutron time of flight
instruments from pinhole- to imaging  configuration in time, thus
enhancing intensity and/or time resolution. New fields of
application for high resolution spectrometry may be opened.
\end{abstract}


\maketitle

\section{Introduction}

A standard lens will image all points $\mathbf{r}_{0}$ from the
object plane to all points $\mathbf{r}_{d}$ in the image plane, and
all rays emerging
from $\mathbf{r}_{0}$ within a certain aperture will arrive at $\mathbf{r}%
_{d}$ no matter what their initial propagation direction was. For
optics in
time one should form the image of an initial event at $t=0\,,\,\mathbf{r}%
_{0} $ at a final event $t_{d}\,,\,\mathbf{r}_{d}$, and to have an
ideal time dependent optical device, all rays leaving the initial
event whatever their initial velocities are, will arrive at the
final event.

Intensive work on electromagnetic imaging in time started in the
late 60's by the development of chirp radar, where pulse
compression\ - imaging in time - in a dispersive medium is achieved
by proper frequency modulation of quadratic shape around the center
of a traveling wave. Here we only refer to the beautiful review
article on temporal imaging by B. Kolner\cite{Ko}, who fully worked
out the space-time duality in imaging, based on the relations
between the Helmholtz equation for paraxial imaging and narrow band
dispersion. Time microscopes and -telescopes are introduced as well
in this paper. More recent work deals with time-dependent
dielectrics in interferometry\cite{Ne}.

In atom optics, time lenses have been proposed and realized. The
rather low matter wave frequencies of ultra-cold atoms can be shifted
by high frequency modulated light, such that matter wave dispersion
in free space leads to temporal imaging. Here we refer to a paper by
A. Arnd et al.\cite{Ar} who observed temporal imaging of atoms via a
light mirror, modulated by a time function which resembles the
spatial function of a Fresnel lens used for spatial imaging. A time
lens for atoms based on a magnetic field of parabolic shape and very
short pulse width was realized by E. Mar\'{e}chal et al.\cite{Ma}.
Time interferometry was realized for cold atoms\cite{Be}, in
shifting their matter wave frequencies by time modulated light
waves.

Longitudinal compression of electron beams to enhance the
performance of free electron lasers has found strong theoretical
interest (see for example \cite{Ro}).

For neutrons, optics in time was stimulated by early papers\cite{Mo}, \cite%
{Ge}, calculating the interaction of a plane wave with an aperture,
which opens suddenly in time. Imaging in time was achieved by high
frequency
mechanical chopping of the amplitude of ultra-cold neutrons\cite{Fr1},\cite%
{Fr3}. The slit pattern on the chopper resembled a one dimensional
Fresnel zone plate, so the modulation frequency changed
quadratically with the time
distance from the 'optical time axis'. Theoretical work on these subjects%
\cite{Fr2} and also on time interferometry\cite{Fe} has been
published.
Closely related phenomena, like diffraction from vibrating surfaces\cite{Fe2}%
, diffraction from time-dependent slits\cite{Hi} and interferences
induced by time-dependent B-fields\cite{Gr} have been observed as
well.

Following the techniques known in charged particle accelerator
physics, a neutron lens can also be created by electromagnetic
forces traveling with the particle beam. Instead of the mainly
electrical forces providing acceleration of charged particles, the
gradient of a magnetic field can be the driving force of a neutron,
possessing a magnetic moment. In agreement with Liouville's theorem
the phase space volume is conserved but the neutrons can be
concentrated into certain intervals. These ideas were first
discussed by H. Rauch and coworkers \cite{Ra},\cite{Su}, who studied
moving magnetic fields with parabolic shape in direction of neutron
propagation. Their aim was to tailor neutron beams supplied by
future pulsed neutron sources. Beam monochromatization i.e. temporal
imaging to infinity, bunching and cooling of polarized neutron beams
were discussed and these methods may advance neutron spectroscopy.

\begin{figure}[tl]
\includegraphics[width=0.46\textwidth]{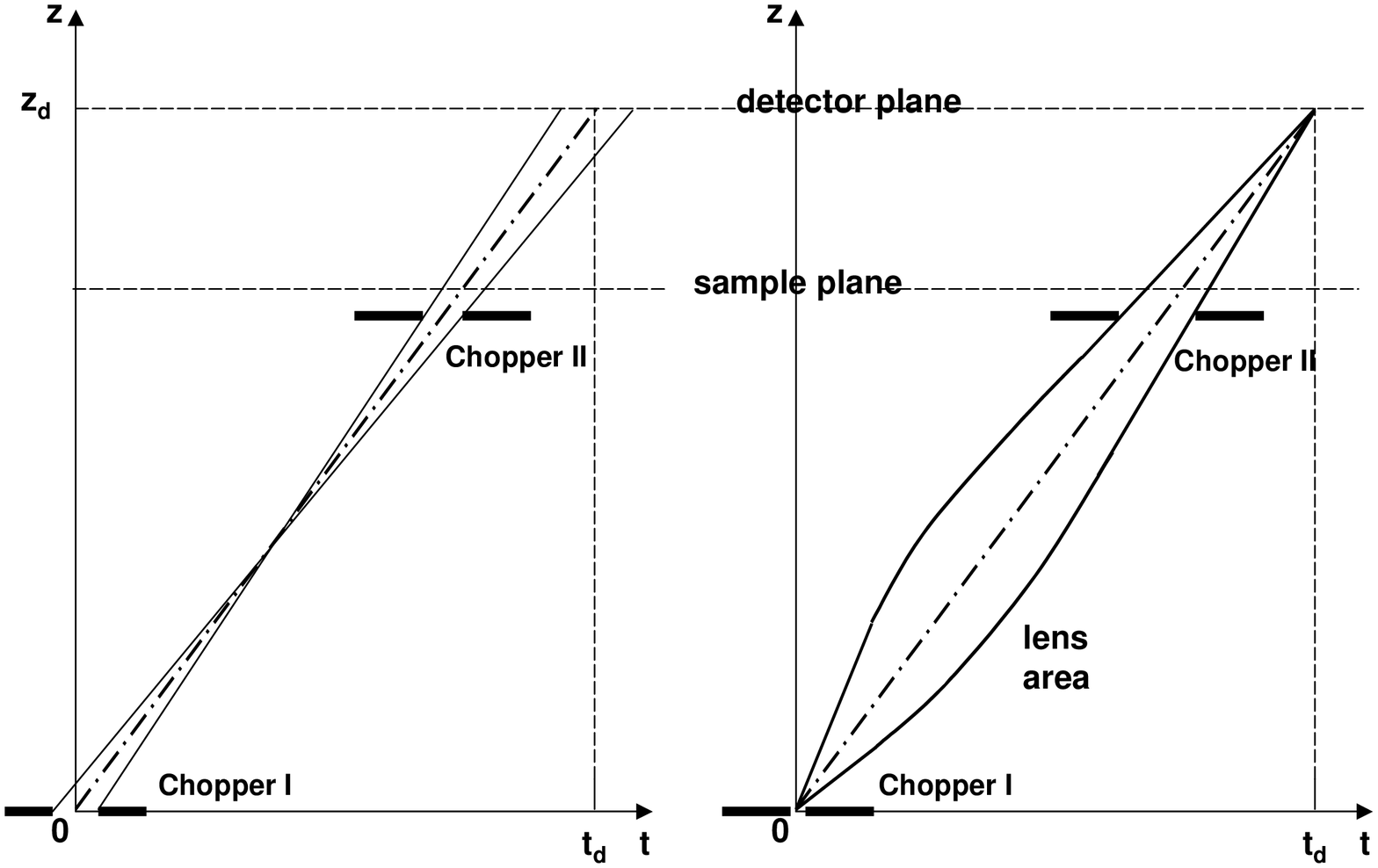}
 \caption{\label{Fig.1} 
 Space-
time diagrams of a standard TOF spectrometer (left) and an imaging
TOF spectrometer (right). In the standard case, the second chopper
acts as a pin hole camera in time, imaging the first chopper opening
time. In the right case, the second chopper merely acts as an
aperture for the time lens.}
\end{figure}

Our proposal is closely related to their idea, but we follow a
different approach and the realization in our case looks less
challenging. We want to replace the pin hole like time optics of
neutron time of flight (TOF) spectrometers by an imaging system
based on a traveling magnetic field. The idea can best be seen from
Fig.\ref{Fig.1}.

In the standard pin hole case, the time width of the pulse at the
detector is a convolution of both chopper openings. In the imaging
case the entrance slit, which may have a rather small time width, is
imaged in time onto the plane of detection and the second slit plays
merely the role of the lens aperture, limiting the effect of lens
aberrations and limiting the spectral width of the pulse. The
imaging system should enhance the intensity and/or the time
resolution for time of flight (TOF) spectrometers and open new
beam-handling capabilities.

Those instruments gather important insights in numerous domains of
present
solid state research, like in the dynamics of high Tc superconductors\cite%
{Te},\cite{La}, in the change of dynamics upon crystal-glass transitions\cite%
{Al}, or in the spin Hamiltonians of magnetic clusters\cite{Ca}. A
high resolution cold neutron TOF spectrometer - the type, we aim to
improve - was one of the first instruments, being formally approved
at the SNS in Oakridge.

The remainder of our paper has the following structure. The next
section contains all basic methodical details and equations
necessary for our investigation. Many of the points made below can
be found in the literature but to our knowledge they have never
being concisely written down aimed to propose and to study a time
lens for neutron TOF spectrometers. Armed with this knowledge we
then discuss possible design parameters and miscellaneous subjects
related to time lens focusing. We close with numerical calculations
for a time lens with realistic parameters.

\section{Calculation of the magnetic field profile}

\begin{figure}
\includegraphics[width=0.5\textwidth]{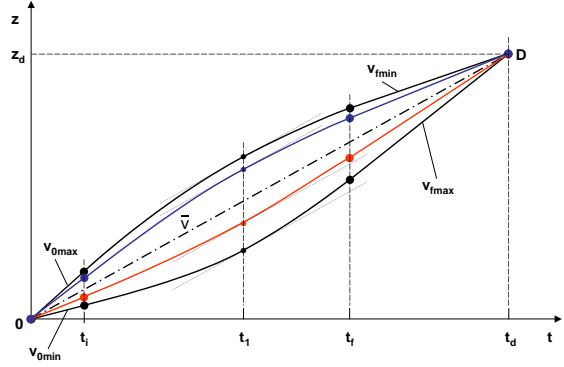}
 \caption{\label{Fig.2} (Color online) Space time diagram for
time imaging. The origin $(z=0;t=0)$\ is imaged to a point
$(z=z_{d};t=t_{d})$\ via a time lens, which action is switched on at
$t_{i}$\ and which is terminated at $t_{f}.$\ At the principal time
$t_{1}$\ all velocities are equal.}
\end{figure}

We want to focus a neutron beam in time, emitted at point $O$:
$t=0,\ z=0$ to point $D$: $t=t_{d},\ z=z_{d}$, (see the
time-coordinate ($t$-$z$) diagram Fig.\ref{Fig.2}), assuming a
paraxial beam along $0\leq z\leq z_{d}$ . For this purpose we need
to apply some external force $F(t,z)$ to the neutrons. This force is
turned on at time $t_{i}>0$ and is turned off at time $t_{f}>t_{i}$.
Before $t_{i}$ or after $t_{f}$ the neutrons move with constant
velocities. To find the required force $F(t,z)$ we assume that the
time interval of emission around $t=0$ is much less than the time
$t_{i}$. Then the trajectories of neutrons do not intersect and are
determined only
by the initial velocities of neutrons $v_{0},$ shown as different slopes in 
the $z$-$t$ diagram. The neutron coordinates at time $t_{i},$ when $F$ is
turned on, are given by
\begin{equation}
z_{i}=v_{0}t_{i}.  \label{z0}
\end{equation}

It is possible to focus the neutrons to the common point $D$ keeping
the trajectories non-intersected. At the time $t_{f}$, when $F$ is
turned off, each neutron must have a velocity $v_{f}$ related to its
coordinate $z_{f}$ given by the equation
\begin{equation}
v_{f}=(z_{d}-z_{f})/(t_{d}-t_{f}).  \label{vf}
\end{equation}
In this case all neutrons will meet at $D$, the space-time focal
point of the system.

The force $F(z,t)$ determines the neutron acceleration $a(z,t)\equiv
F(z,t)/m_{n}$, where $m_{n}$ is the neutron mass. We will look for a
special solution $a(z,t)$, where for each neutron $j$ the
acceleration will be constant during the full time range from
$t_{i}$ to $t_{f}$. For each neutron the acceleration is only
determined by its initial and final velocities via \ \
\begin{equation}
a_{j}=(v_{f}-v_{0})/(t_{f}-t_{i}).  \label{a1}
\end{equation}%
With this choice, the acceleration of each neutron is constant in
time but
depends on the initial velocity.

\begin{figure}
\includegraphics[width=0.46\textwidth]{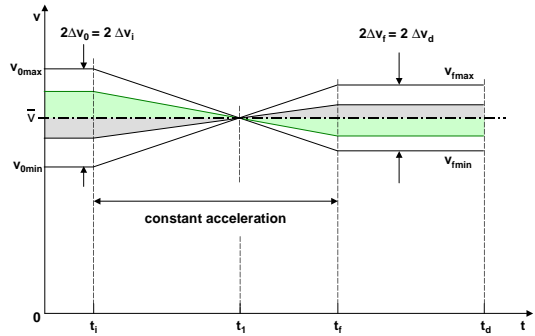}
 \caption{\label{Fig.3} (Color online) The integrals (shaded areas) 
 for each individual particle velocity
left and right from the principal time $t_{1}$ must be equal in
order to fulfill the imaging conditions.}
\end{figure}

The above conditions, together with the non-intersection of all
trajectories $z(v,t)$ and the constant acceleration between $t_{i}$
and $t_{f}$ for each neutron, imply a velocity-time diagram as shown
in Fig.\ref{Fig.3}, where at a certain time $ t_{1}$ all velocities
$v$\ have the same values $v=\overline{v}=$\ $ z_{d}/t_{d}.$ Note
that for all trajectories (some are shown in Fig.\ref{Fig.3}) the
shaded areas left and right from $t_{1}$\ must be equal, as
\begin{equation}
z_{d}=\int_{0}^{t_{d}}v(t)dt  \label{zd}
\end{equation}
is equal for all neutrons. In analogy to optics in space, $t_{1}$ is
called the principal time of the system. From Fig.\ref{Fig.3} and
Eq. (\ref {zd}) we get:
\begin{equation}
t_{1}\equiv t_{f}-\frac{t_{f}^{2}-t_{i}^{2}}{2t_{d}}  \label{t1}
\end{equation}%
Imaging to infinity in time $(t_{d}\rightarrow \infty )$, requires $
t_{1}=t_{f}$, i.e. the lens action is terminated at the principal
time, when all neutrons have the same velocity.

The velocity-time diagram Fig.\ref{Fig.3} also shows that the
velocity bandwidth changes from $2\Delta v_{0}$ at $t_{i}$ to
$2\Delta v_{d}$ at $ t_{d}$ with
\begin{equation}
\Delta v_{0}/\Delta v_{d}=\left( t_{1}-t_{i}\right) /\left(
t_{f}-t_{1}\right)  \label{dv}
\end{equation}

\begin{figure}
\includegraphics[width=0.48\textwidth]{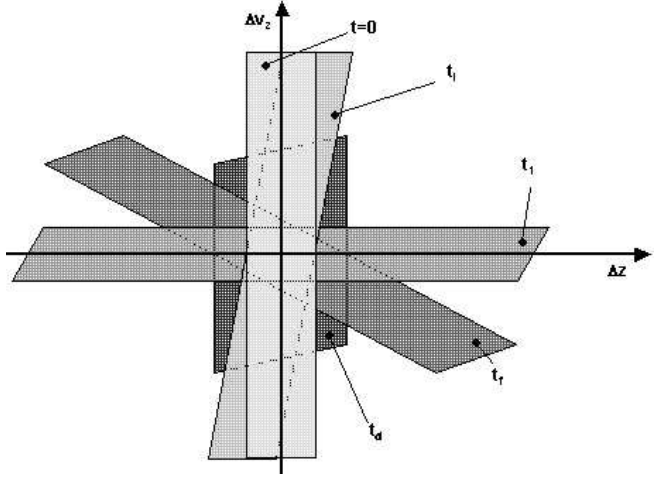}
 \caption{\label{Fig.4} The evolution of the
phase space element with time $(t=0\rightarrow t_{i}\rightarrow
t_{1}\rightarrow t_{f}\rightarrow t_{d})$ calculated from the
evolution of its four corner points. From this picture, a simple
formula for the time magnification $M$ is obtained: $M=\Delta
t_{d}/\Delta t_{0}$ $=\Delta v_{0}/\Delta v_{d}.$ In our case, the
time magnification is $M\approx 2.$}
\end{figure}

The evolution of a small rectangular phase space
element with time upon the passage of the pulse through the optics
(obtained analytically and then checked by the Monte Carlo calculation) 
is presented in Fig.\ref{Fig.4}.  
It shows (for details see ch. 4) that for small enough values of
$\Delta v_{0}$ and $\Delta z_{0}$ a rectangular phase space element at $t_{0}$\
transforms to a parallelogram (near rectangular) phase space element 
at $t_{d}$. Due
to Liouville's theorem the phase space volume of the neutron pulse
remains constant \cite{comment1} - no 'brightening' of the focus
occurs due to the time-dependent magnetic field \cite{Ke}. Hence, a
simple estimate for the time magnification $M$ can be obtained,
where $M$ is defined as the ratio $\Delta t_{d}/\Delta t_{0}$ of
pulse widths at $t=t_{d}$ and $t=0$:
\begin{equation}
M=\Delta t_{d}/\Delta t_{0}\approx \Delta z_{d}/\Delta z_{0}=\left(
t_{1}-t_{i}\right) /\left( t_{f}-t_{1}\right) .  \label{Ma}
\end{equation}

We now derive the magnetic field in the interval between $t_{i}$ and
$t_{f}.$ The coordinate of a neutron at time $t_{f}$ is given by
\begin{equation}
z_{f}=v_{0}t_{f}+a(t_{f}-t_{i})^{2}/2.  \label{zf}
\end{equation}
By substituting (\ref{vf}) and (\ref{zf}) into (\ref{a1}) we get a
relation between $a$ and $v_{0}$:
\begin{equation}
a=\frac{z_{d}-v_{0}t_{d}-a(t_{f}-t_{i})^{2}/2}{(t_{f}-t_{i})(t_{d}-t_{f})}
\label{a2}
\end{equation}
which rewrites as
\begin{equation}
a(v_{0})=\frac{2(z_{d}-v_{0}t_{d})}{(t_{f}-t_{i})(2t_{d}-t_{f}-t_{i})}=\frac{%
\overline{v}-v_{0}}{t_{1}-t_{i}}  \label{a3}
\end{equation}
where $\overline{v}=z_{d}/t_{d}$ is the central velocity of the
neutrons,
which is not affected by the lens action $[a(\overline{v})=0]$. Eq. (\ref{a3}%
) determines the acceleration as function of the initial velocity
$v_{0}$ of neutrons, it can easily be verified by looking at
Fig.\ref{Fig.3}. With this acceleration each neutron will come to
the point $z_{d}$ at time $t_{d}$. To find this acceleration as
function of time and coordinate $a(t,z)$ we need to use the relation
between the initial velocity of a neutron, its coordinate and time.
In the interval $t\in \lbrack t_{i},t_{f}]$ this relation is
\begin{equation}
z(t)=v_{0}t+a(t-t_{i})^{2}/2.  \label{zt}
\end{equation}
Substituting it into (\ref{a3}) we obtain a linear equation on $a$:
\[
a=\frac{\overline{v}-[z(t)-a(t-t_{i})^{2}/2]/t}{(t_{1}-t_{i})}
\]%
and solving for $a(t,z)$ we get
\begin{equation}
a(t,z)=\frac{2(\overline{v}t-z)}{t_{1}^{2}-t_{i}^{2}-(t-t_{1})^{2}}.
\label{a4}
\end{equation}
The required external force is given by $F(t,z)=m_{n}a(t,z)$. If
this force is due to the gradient of the magnetic field which acts
on spins of polarized neutrons, the required magnetic field as
function of $(z,t)$ is determined by
\begin{equation}
\mu _{n}\frac{\partial B}{\partial z}=-m_{n}a(t,z),  \label{dB}
\end{equation}
where $\mu _{n}$ is the magnetic moment of neutron. Writing this
relation we assumed that the time derivative of the magnetic field
is negligible compared to its space derivative:
\[
\left\vert \frac{\partial B(t,z)}{\partial z}\right\vert \gg \frac{1}{c}%
\left\vert \frac{\partial B(t,z)}{\partial t}\right\vert ,
\]%
where $c$ is light velocity. After integration of (\ref{a4}) over
$z$ we obtain
\begin{equation}
B(t,z)=\frac{m_{n}}{\mu _{n}}\frac{(z-\overline{v}t)^{2}}{%
t_{1}^{2}-t_{i}^{2}-(t-t_{1})^{2}}  \label{B}
\end{equation}
This magnetic field profile is just a parabola widening and
shrinking in time and moving with the central velocity
$\overline{v}$ of neutrons. The curvature of the parabola changes
with time as determined by the denominator in formula (\ref{B}).

The sign of the field is spin dependent, so that the desired
focusing effect will occur for neutrons with one of the two
polarizations, what is also significant for aberrations as we
shortly discuss below.

The condition of constant acceleration for each neutron implies that
each one remains at a position of constant slope of the parabola,
and the spatial extension of the neutron pulse in z direction is
equal to the width at
maximal slope of the parabola, necessary to focus the initial velocity band $%
\Delta v_{0}$. The maximal width of the parabola is reached at
$t=t_{1},$ (the denominator of Eq. (\ref{B}) is maximal there). At
this time the spread of the beam is maximal too and all neutrons are
at rest at $t_{1}$\ in the frame moving with $\overline{v},$ (see
also Fig.\ref{Fig.3}).

In the moving frame ($\overline{v}$) we calculate the width
$w(\Delta v)$ of
the magnetic bowl. With $v_{0}=\overline{v}+\Delta v_{0}$ we get from (\ref%
{a3}):
\begin{equation}
a(\Delta v_{0})=-\frac{\Delta v_{0}}{t_{1}-t_{i}}  \label{ad}
\end{equation}
which we set equal to Eq. (\ref{a4}) with $z=\overline{v}t+w,$
leading to:
\begin{equation}
2w(t)=\Delta
v_{0}\frac{t_{1}^{2}-t_{i}^{2}-(t-t_{1})^{2}}{(t_{1}-t_{i})}\;
\label{wt}
\end{equation}
Replacing $z$ by $w$ in Eq. (14) we get for the B-field in the moving frame:
\begin{equation}
B(t,\Delta v_{0})=\frac{m_{n}}{\mu _{n}}\frac{\Delta v_{0}^{2}}{4}\ \frac{ 
t_{1}^{2}-t_{i}^{2}-(t-t_{1})^{2}}{(t_{1}-t_{i})^{2}}  \label{Bd}
\end{equation}

\begin{figure}
\includegraphics[width=0.48\textwidth]{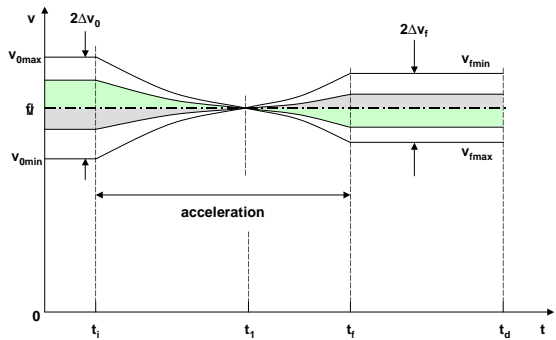}
 \caption{\label{Fig.5} (Color online) Velocity-time diagram of a
time lens with varying acceleration. In this case the maximum field
at $t_{1}$\ is reduced compared to the case of constant
acceleration. As before the shaded areas left and right from the
principal time $t_{1}$\ must be equal.}
\end{figure}

The condition of constant acceleration, assumed at the beginning, is
not mandatory to achieve focusing in time. 
A non-linear velocity
curve as shown in Fig.\ref{Fig.5} can be used as well, as long as
(\ref{zd}) is fulfilled. In this case the maximum field strength may
be significantly reduced, as the acceleration at $t_{1}$, where $w$
is maximum, is lower now. However, in this case the field rise and drop near 
$t_0$ and $t_f$ gets faster, and the adiabaticity condition
(the neutrons must smoothly turn to the direction of the $B$-field) may be more
difficult to fulfill. So it may turn out
that low maximal fields at $t_0$ and $t_f$ are more favorable.
These questions request numerical evaluation, which have
not yet been done.

Aberrations of magnetic time lenses may arise from field gradients
in directions lateral to the optical axis. They may be analyzed in
the moving frame, considering $B(t,z)$ as slowly varying function
with respect to $t$.

Near the optical axis, from $\nabla \cdot B=0$ and axial symmetry with $ 
\dfrac{\partial B_{x}}{\partial x}=\dfrac{\partial B_{y}}{\partial
y}$ we
get $\dfrac{\partial B_{z}}{\partial z}=-2\dfrac{\partial B_{x}}{\partial x}$%
, leading to lateral accelerations $a_{x},a_{y\text{ }}$ which,
according to
Eq. (\ref{ad}) are given by:
\[
a_{x}(\Delta v_{0})=a_{y}(\Delta v_{0})=\frac{\Delta
v_{0}}{2(t_{1}-t_{i})}
\]
The characteristic dimensions of the beam cross section, defined by
neutron guides, will be very small compared to the length of the
time lens and consequently the path lengths for acceleration in
lateral direction are very
small. For reasonable parameters, the mean change in the lateral velocities $ 
v_{x},v_{y\text{ }}$picked up due to lateral accelerations will be
in the range of $\pm 2m/s$ (see below). This is still small compared
to typical values $v_{x},v_{y\text{ }}$, determined by the guide
properties ($\simeq \pm 10m/s$) and no significant beam widening
should occur.

Further aberrations arise for off-axial neutrons, as there the acceleration $ 
a_{z}(\Delta v_{0})$ depends on the distance from the optical axis.
Due to
the beam divergence of typically $1^{\circ },$ averaging over the different $ 
a_{z}(\Delta v_{0})$ will occur. We analyzed these effects
numerically, as they depend in a complex way on the beam geometry
and\ on the shape of the coils which generate the B-fields. No
serious distortion of the time signal was found.

One more point for the sake of a skeptical reader remains to be
clarified. As it is well known in the classical space optics, it is
impossible to obtain a geometrically similar image even for an
infinitely small object (except of the trivial case of identical
infinite plane mirror imaging). The fact is a trivial consequence of
that the longitudinal magnification is never equal to the transverse
one. However since our aim is not to have an ideal image of an
initial $t_{i},$ $r_{i}$ event but to compress a neutron beam to
enhance the intensity for the TOF devices in a certain velocity -
time interval, this kind of aberrations, which is a nuisance for
space optics, for our aims on the contrary provides valued effects
improving TOF time resolution.

There is another cause of aberrations in the standard space optics.
It is related to paraxial or Gaussian approximations which are based
on an expansion of phase factors over off-axis distances\cite{BW}.
Formally we have found the exact space - time trajectory Eq.
(\ref{zt}), Eq. (\ref{a4}), or by other words all aberrations are
included (the validity window for the geometrical optics approach is
extremely wide since thermal neutron wavelengths are of the order of
$1\mathring{A}$ much smaller than typical sample sizes).

Thus, in principle, in the frame work of the found above special \
solution of the Newton equations, an arbitrary good resolution can
be achieved (except that evident physical and technical constraints
we discuss in the next section). For a more general case (when the
condition of constant acceleration for each neutron is not granted),
the analytical treatment requires a set of different approximate
methods on certain scales of length and time, and one has to
consider the aberration problems.

\section{Design parameters and gain factors}

For efficient use in a TOF spectrometer, the magnetic lens should be
located between the first chopper and the sample (see
Fig.\ref{Fig.1}). The lens region should be as long as possible to
keep the maximum field strength $B$\ for a given velocity band
$\Delta v$ as low as possible. In TOF spectrometers, in general the
distance $L_{p}$ from the first chopper to the sample is somewhat
larger than the distance $L_{s}$ from the sample to the plane of
detection, which helps for choosing a time magnification $M$ close
to 1 (see Eq.(\ref{Ma})) with its optimal imaging properties
concerning time resolution at the detector.

As design parameters we choose $L_{p}=14m$, $L_{s}=4m$ and a total
length of $L_{B}=8m$ for the traveling field, starting $5m$ behind
the first chopper and terminating 1m before the sample position. As
central neutron velocity we take $\overline{v}=600m/s,$which is a
convenient velocity for high resolution TOF instruments. From these
parameters we obtain for the central
neutron velocity: $t_{i}=8.3ms;$ $t_{f}=21.7ms;$ $t_{d}=30ms;$ $t_{1}=15ms;$
(see Eq. (\ref{t1})) and $M=1$ (see Eq. (\ref{Ma})).

As minimal opening time $2\Delta t_{0}$ of the first chopper we take $%
2\Delta t_{0}=5\mu s$, which is reached by state of the art
double-chopper systems for typical beam cross sections of 3 cm
width. With $\Delta t_{d}=M\cdot \Delta t_{0}$, a minimum pulse
width at the detector of $\ 2\Delta t_{d}=5\mu s$ is achieved. This
still matches to the time resolution of commonly used gas detector
tubes, however the use of scintillator based detectors (at least 10
times higher resolution) seems preferable in this case.

Eq. (\ref{Bd}) links the maximal velocity band $\Delta v_{0}$,\
which can be focused by the time lens, to the maximum traveling
$B-$field, which can be applied. Due to the rather low duty cycle of
less than 10\% of the field, a maximum field strength of $B=1T$ can be
assumed in the following, certainly
within reach using state of the art accelerator technologies. From Eq. (\ref%
{Bd}) with $t=t_{1}$ (where $B$ reaches its maximum value) we get
\begin{equation}
\Delta v_{0}^{2}=4\frac{B(t_{1},\Delta v_{0})}{m_{n}/\mu _{n}}\ \frac{%
t_{1}-t_{i}}{t_{1}+t_{i}}  \label{dv2}
\end{equation}%
For the above parameters and $m_{n}/\mu _{n}=0.173\ kgT/J$ we get
for the velocity band $2\Delta v_{0}$ to be focused by the time
lens:

\[
2\Delta v_{0}=5.1m/s
\]

Note that this bandwidth depends on the geometry of the spectrometer
but to first order it does not depend on the neutron velocity, as
the second factor does not depend on $v$. This favors slow neutrons,
as the relative bandwidth increases with wavelength.

A second chopper should be placed at a position
$L_{1}=\overline{v}t_{1};$ it takes the role of the time aperture
for the optics, limiting the velocity band to the nominal width
$2\Delta v_{0}.$ Actually it is more convenient to place this
chopper after the wanderfeld area with slightly reduced opening
time.

For $t_{1}\gg t_{i}$, i.e. when the neutrons enter the B-field very
close
after the first chopper, Eq. (\ref{dv2}) gives:%
\begin{equation}
\frac{1}{4}m_{n}\Delta v_{0}^{2}=\mu _{n}B(t_{1},\Delta v_{0})
\label{energ}
\end{equation}%
This manifests energy conservation in the moving frame, stating that
the neutron climbs up the wall of the magnetic bowl, which is very
low at $t_{i}$ and maximal at $t_{1}.$ The unusual pre-factor 1/4
can easily be understood: in a parabolic bowl, which is constant in
time, the force increases linearly with the distance from the
center, in our case however the force is maximal and constant due to
the assumption of constant acceleration. Integrating the force over
the distance gives in a factor of 2 more in our case compared to the
constant bowl.

In the lab-frame, the energy change $\Delta E$ of the neutron is
given by:
\[
\Delta E=mv_{0}\cdot \Delta v_{0}
\]%
which is about 2 orders of magnitude higher than in the moving frame, as $%
v_{0}\approx 100\cdot \Delta v_{0}$. This demonstrates the high
efficiency of the wanderfeld focusing technique.

As usual, there is no unequivocal way to define gain factors. We
will compare the present setup with a reference TOF spectrometer,
showing the same nominal time resolution of $2\Delta t_{d}=5\mu s$\
at the detector,
based on a geometry, which is better adapted for that case: We choose $%
L_{p}^{\prime }=6m$, $L_{s}=4m$, with the second chopper in a
distance of 5m after the first one, i.e. centered between the first
chopper and the detector. In this geometry the nominal time
resolution of $5\mu s$ is obtained for $2\Delta t_{0}^{\prime
}=\sqrt{2}\cdot 5\mu s\ $and $2\Delta t_{1}^{\prime
}=1/\sqrt{2}\cdot 5\mu s$ for the first and second chopper
respectively, neglecting further broadening due to flight path
uncertainties. The gain $G$ in intensity by the wanderfeld technique
can be estimated by:
\begin{equation}
G=0.4\cdot \frac{\Delta t_{0}}{\Delta t_{0}^{\prime }}\frac{\Delta v_{0}}{%
\Delta v_{0}^{\prime }}  \label{gain}
\end{equation}%
where $2\Delta v_{0}^{\prime }$ is the velocity bandwidth
transmitted by the second chopper of the reference spectrometer, in
case the time opening of the first chopper would be infinitely
small. The factor 0.4 takes into account the loss in intensity due
to beam polarization, necessary for the wanderfeld technique. It
includes 20\% loss due to non-ideal transmission of the polarizer.
Polarizations of 98-99\% are easily in reach with state of the art
super mirror polarizers or long polarizing guides. For the above
parameters we get $2\Delta v_{0}^{\prime }=0.26m/s$ and for the gain
factor G we get:

\[
G=5.5
\]

This gain is proportional to$\sqrt{B},$ and to $1/\overline{v}.$
With more favorable values than taken for the estimate above, gains
of one order of
magnitude seem within reach. Furthermore the envisaged time resolution of $%
2\Delta t_{d}=5\mu s$\ at the detector seems not reachable for
conventional TOF spectrometers, as a value of $2\Delta t_{1}^{\prime
}=1/\sqrt{2}\cdot 5\mu s$ for the opening time of the second chopper
cannot be reached with state of the art technology for a neutron
beam of reasonable width. We are convinced that the time lens will
improve significantly the time resolution in TOF.

\section{Numerical Calculations}

We accompanied the analytic approach by Monte Carlo simulations for
(i) the evolution of the phase space element during propagation
through the optical system; (ii) the dependence of the time
resolution on the energy transfer at the sample; (iii) the effect of
sample size on time resolution; (iv) the optimization of the coil
geometry, which generates the traveling field.

\begin{figure}
\includegraphics[width=0.45\textwidth]{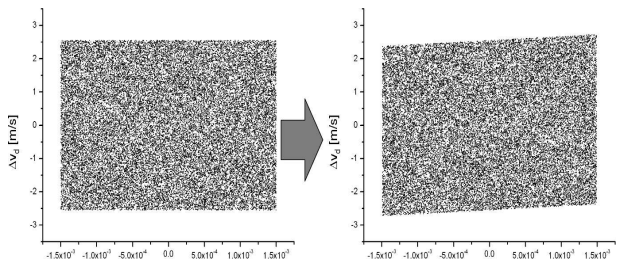}
 \caption{\label{Fig.6} The evolution of phase
space element for the ideal time lens with time magnification $M=1$
from the start to the detector, calculated for about $3\cdot 10^{4}$
neutrons. The density within the volume remains constant.}
\end{figure}

\begin{figure}
\includegraphics[width=0.23\textwidth]{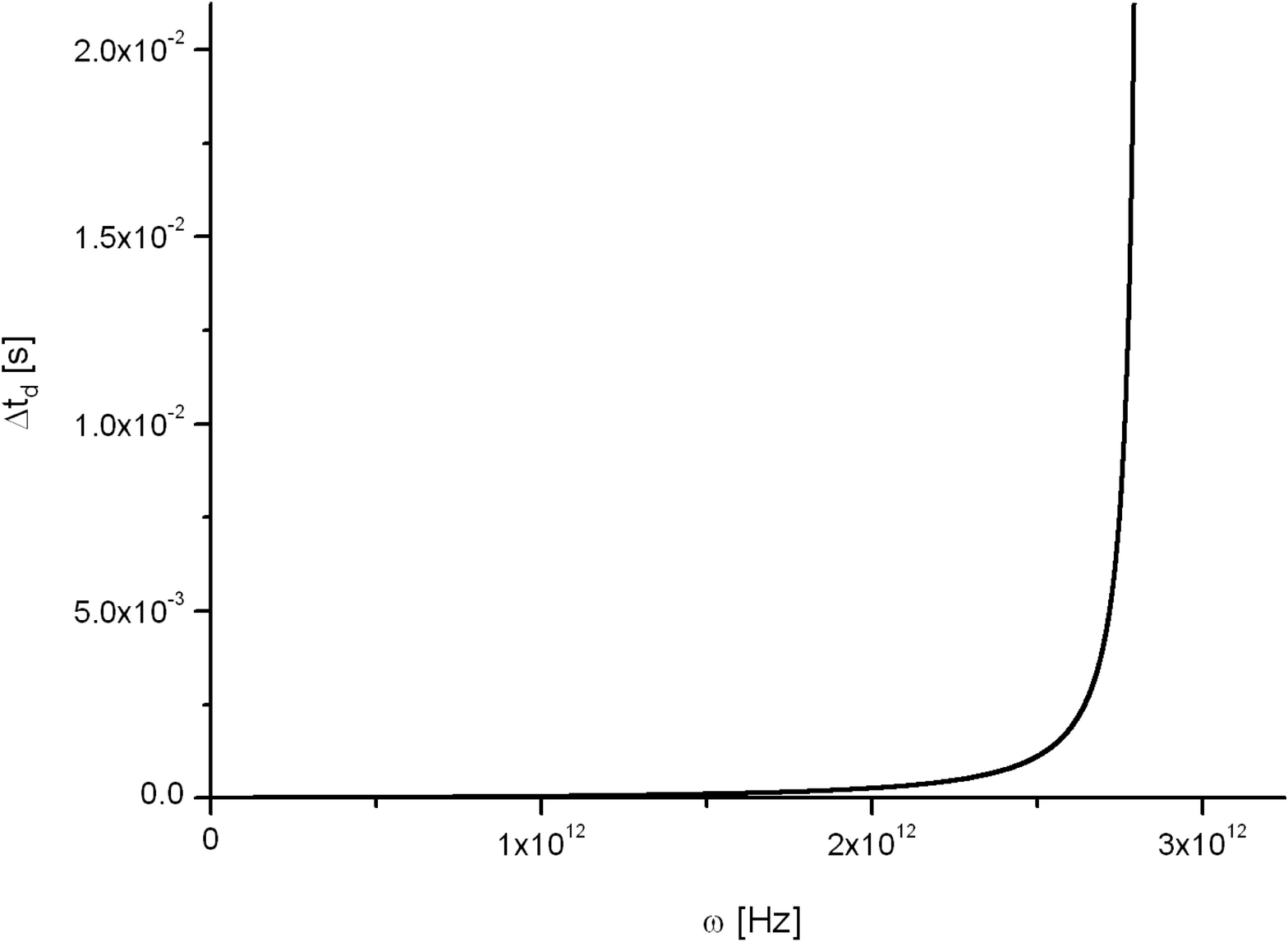}
\hspace{0.05cm}
\includegraphics[width=0.23\textwidth]{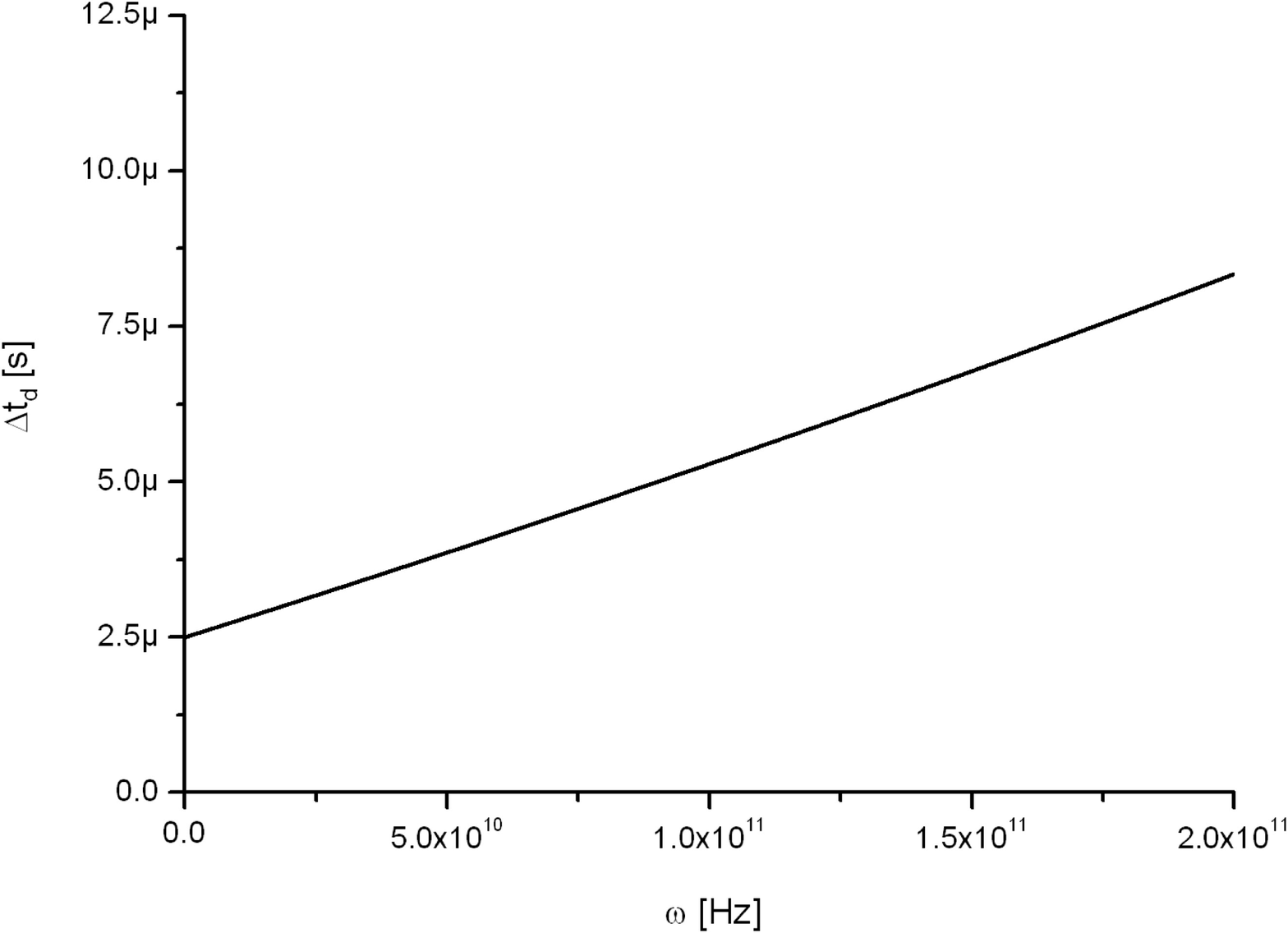}
 \caption{\label{Fig.7} Evolution of the time
resolution $\Delta t_{d}$ as function of the energy transfer
$\hbar \omega $ upon scattering. The nominal matter wave
energy is $\hbar \omega_{0}=12 meV$ (on the right panel 
the linear part of the time resolution dependence is shown).}
\end{figure}

\begin{figure}
\includegraphics[width=0.22\textwidth]{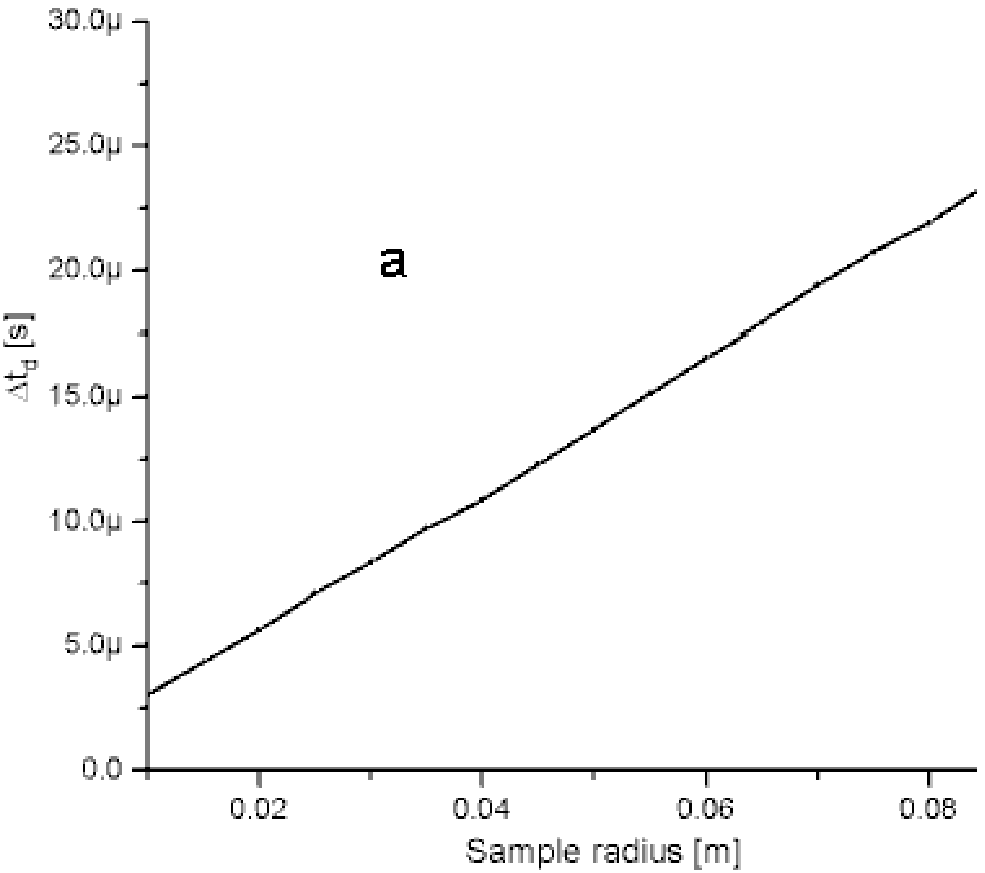}
\hspace{0.4cm}
\includegraphics[width=0.22\textwidth]{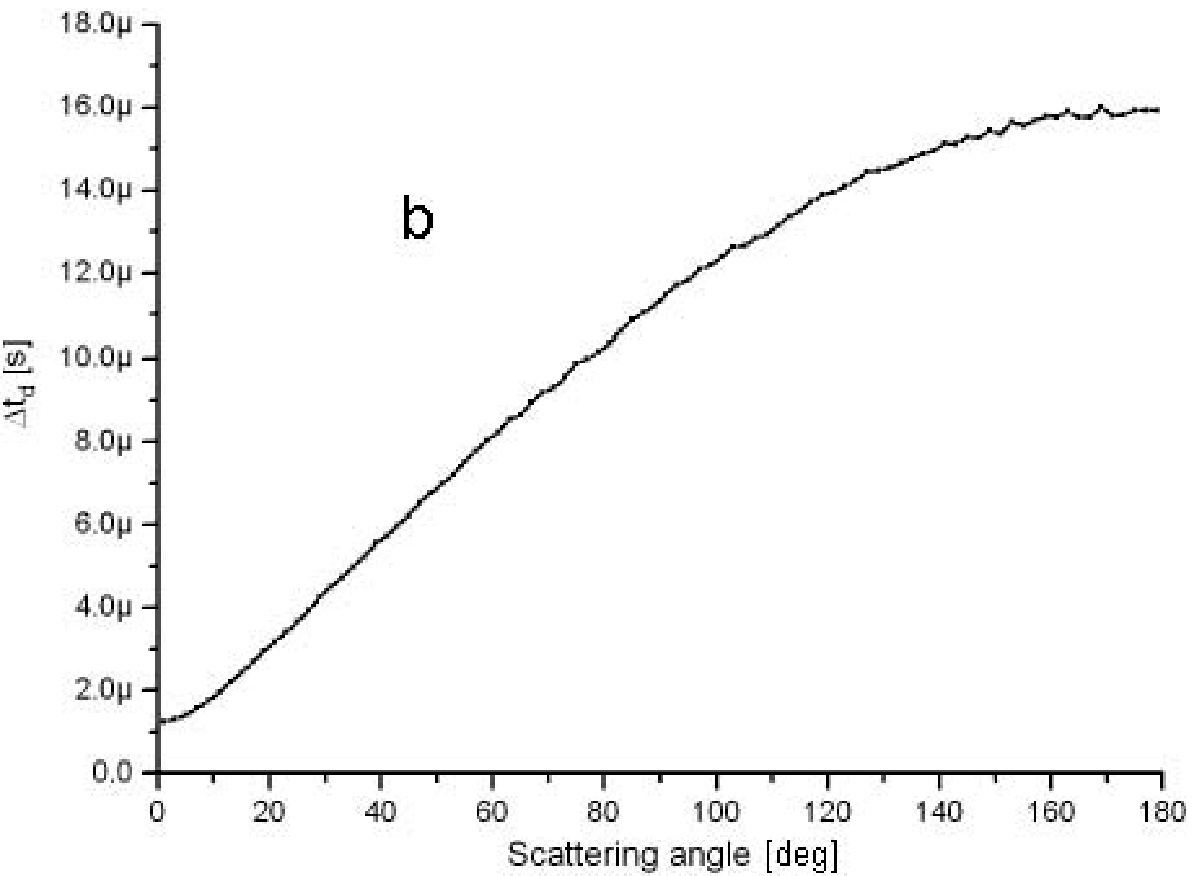}
 \caption{\label{Fig.8} Effect of sample size
and scattering angle on time resolution for a circular sample:
the left panel (a) shows that the time width increases linearly with the radius of the
sample for a specific scattering angle (45${}^\circ $ in the
figure); the right panel (b) shows the dependence of the time width on the
scattering angle (for 1.5 cm sample radius). The fluctuations
originate from statistics. }
\end{figure}

\begin{figure}
\includegraphics[width=0.48\textwidth]{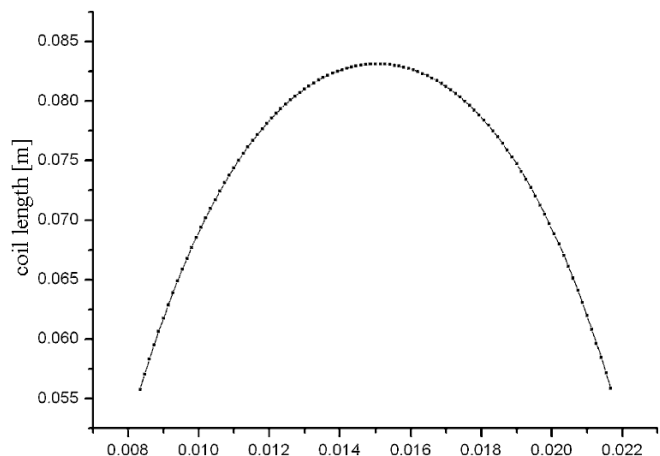}
 \caption{\label{Fig.9} Variation of coil length from
the start time $t_{i}$ of the time lens to the end time $t_{f}$. Its
maximal length is at the principal time $t_{1},$ where the neutron
pulse reaches its maximal length. The other parameters are given in
the text.}
\end{figure}

\begin{figure}
\includegraphics[width=0.48\textwidth]{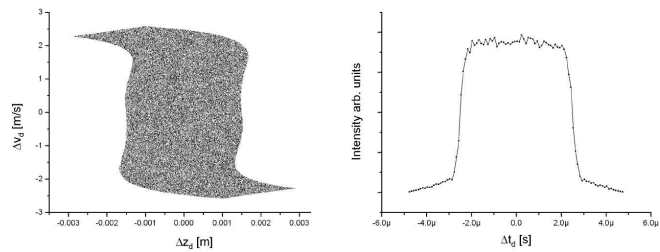}
 \caption{\label{Fig.10} Phase space element
and time signal at the detector for a realistic setup of coils,
generating the traveling field. The 'wings' on both figures are due
to a slight deviation from parabolic field shape, experienced by
those neutrons, which are at the leading and lagging edge of the
traveling neutron pulse.}
\end{figure}

\begin{figure}
\includegraphics[width=0.48\textwidth]{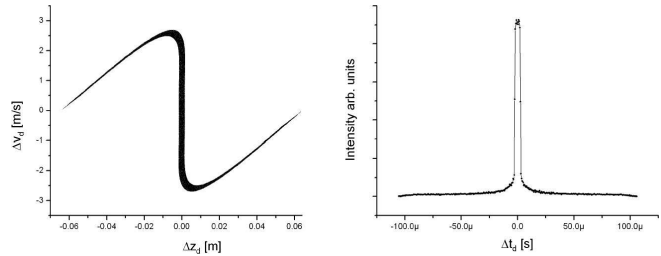}
 \caption{\label{Fig.11} Phase space element
and time signal at the detector for the same traveling field as in
Fig. \ref{Fig.10}, but for a velocity bandwidth nearly twice too large to get
properly focused by the lens. The 'wings' seriously distort the
time resolution.}
\end{figure}

The calculations with the ideal parabolic field (figures \ref{Fig.6},\ref{Fig.7})
 are done in one space and one time coordinates. For the realistic coil design
(figures \ref{Fig.8}-\ref{Fig.11}) a parallel beam of circular cross section with axial 
symmetry is considered.

If not specified differently, we used the following parameters:
\newline mean neutron velocity: $\overline{v}=600m/s;$
mean velocity
spread: $2\Delta v_{0}=5.1m/s;$
mean chopper opening time: $2\Delta
t_{0}=5\mu s;$ distance first chopper to sample: $L_{p}=14m;$
distance sample to plane of detection: $L_{s}=4m;$ total length of
the traveling field: $L_{B}=8m$ starting $5m$ after the first
chopper.

The evolution of the phase space element for the ideal time lens, i.e.
for a perfect quadratic potential, (Eq. (\ref{B})), was already
shown in Fig.\ref {Fig.4}. Using the above parameters,
Fig.\ref{Fig.6} shows the phase space volume at the start and at the
detector, calculated for about $3\cdot 10^{4}$ neutrons; it clearly
demonstrates that the density within the volume remains constant
upon its transform. This also confirms the formula for time
magnification (see Eq. (\ref{Ma})) and apparently Liouville's
theorem holds.

As the TOF- technique is applied for inelastic scattering, we
checked the influence of energy change $\hbar \omega $ upon sample
scattering on the time resolution of the time optics. Up to $\omega
-$values of about $ 2.0\cdot 10^{11}Hz$ ( $\approx 10\%$ of neutron
energy), time resolution stays below $\Delta t_{d}=8\mu s$ (see Fig.
\ref{Fig.7}) . Up to a \ $\omega =2.0\cdot 10^{12}Hz$, corresponding
to about 70\% of the neutron energy, and the time widening of the focus
goes fairly linear with $\omega $.

Next we examined the effect of sample size and scattering angle on
time resolution, assuming a circular sample and homogeneous
scattering within. The time width increases linear with the radius
of the sample for a specific scattering angle as seen on Fig.
\ref{Fig.8}a. The time width starts to increase quadratically with
scattering angle, and finally reaches a maximum of about $\Delta
t_{d}\approx 16\mu s$ at a scattering angle of $\pi $ for a sample
of 1.5 cm radius (Fig. \ref{Fig.8}b).

Next we derived a realistic coil design for a traveling field of
near-parabolic shape and calculated for this field the phase space
element at the detector and the corresponding time resolution. A
magnetic field of axial geometry with parabolic shape in z-direction
implies $\partial B/\partial x=\partial B/\partial y\neq 0$. As
discussed earlier, lateral forces proportional to $\partial
B/\partial x$ or $\partial B/\partial y$ will not disturb
significantly the lens action and therefore were neglected. However
we took into account the reduction of the longitudinal force with
the distance from the optical axis, resulting from $\nabla
B=0$.

We assume the field to be realized by a set of cylindrical coils of
varying shape and field strength along the z-axis. In the
calculation we take a coil traveling with $\overline{v}$ with
varying radius, length and current. The following calculations were
done for a coil radius of 4 cm, a current of
217A, 250 turns/cm and for varying length, as shown in Fig. \ref{Fig.9}.

Since the duty cycle of the coil
is only in the \%-range, the assumed current should be applicable. Fig.\ref%
{Fig.10} shows the phase space element and the time signal at the
detector for the setup in mind.

For a bandwidth $\Delta v_{0}=4.4%
\frac{m}{s}$, ($2.5\frac{m}{s}$was used in the example before), the
phase space element and the time distribution show very pronounced
tails on both sides, as seen in Fig.\ref{Fig.11}. Such a broad
bandwidth clearly is outside of the focusing capacities of the time
lens with the parameters in
mind.

We have demonstrated the feasibility of a time lens, which may open
new fields for high resolution neutron TOF instruments. Further
calculations
will focus on minimizing the tails in the time distribution for larger 
$\Delta v_{0}$.

\acknowledgements

Many thanks are due to Rupp Lechner from HMI-Berlin for his
explanations on TOF resolution and for critical reading of the
manuscript.


\begin{thebibliography}{99}
\bibitem{comment1} Generally, the Liouville's theorem does not hold for time-dependent
Hamiltonians. Luckily, for our special solution (constant
acceleration along a particle trajectory with the parabolic
potential) the theorem does work (see also our numerical results
collected in the section 4).
\end{thebibliography}
\end{document}